\documentclass{article}

\usepackage[utf8]{inputenc}
\usepackage{xcolor}
\usepackage{subfig}
\usepackage{longtable}
\usepackage{multirow}
\usepackage{subfig}

\usepackage{graphicx}
\begin{document}
%

\title{A Methodology to Support Automatic Cyber Risk Assessment Review}
%
%
\date{}
\author{Marco Angelini \and
Silvia Bonomi \and
Alessandro Palma
}
%
%
\maketitle              
\begin{abstract}
Cyber risk assessment is a fundamental activity for enhancing the protection of an organization, identifying and evaluating the exposure to cyber threats.
Currently, this activity is carried out mainly manually and the identification and correct quantification of risks deeply depend on the experience and confidence of the human assessor. As a consequence, the process is not completely objective and two parallel assessments of the same situation may lead to different results.
This paper takes a step in the direction of reducing the degree of subjectivity by proposing a methodology to support risk assessors with an automatic review of the produced assessment.
Our methodology starts from a controls-based assessment performed using well-known cybersecurity frameworks (e.g., ISO 27001, NIST) and maps security controls over infrastructural aspects that can be assessed automatically (e.g., ICT devices, organization policies). Exploiting this mapping, the methodology suggests how to identify controls needing a revision.
%
The approach has been validated through a case study from the healthcare domain and a set of statistical analyses.\\

\end{abstract}
\section{Introduction}
\label{sec:intro}

The growing number of cyber threats puts more and more organizations at risk of being the victim of a cyber attack. In this picture, it is thus paramount to design and implement efficient and effective processes to timely identify and manage potentially dangerous situations.
Cyber risk management is one of those processes, maybe the most important one. It is challenging due to the peculiarities of cyberspace (e.g., wide attack surface and a high degree of uncertainty) and its main activities are identified by several standards and best practices (\cite{ISO31000},\cite{ISO27500},\cite{NIST800-100}).
Evaluating cyber risk raises several issues: human errors are always possible in each step of the assessment, as well as different sensibilities from different assessors to similar situations that may influence the evaluation. 
Those factors could negatively affect the reliability of the final outcome. Moreover, the bias during the assessment could compromise its accuracy.
This paper proposes a methodology 
to review the results of a risk assessment depending on the actual organization's infrastructure exposure.
We introduce the concept of \emph{comprehensive score} to quantify the security of the different parts of the infrastructure at different levels (i.e., technical, human-based) and take into account governance aspects.
To perform this comprehensive analysis, we exploit the multi-layer attack graph (MLAG), over which the results of an assessment based on well-known frameworks (e.g., ISO 27001, NIST) are mapped.
Through this approach, the assessor has more fitting quantification of the cyber risk and the critical exposures in the infrastructure.

\section{Related work}
\label{sec:rel}

In the current literature, there exist several methodologies (and sometimes related tools) developed in compliance (sometimes partially) with the standards, to support the risk identification, analysis, and treatment. 
For example, MEHARI risk management methodology \cite{MEHARI} and EBIOS \cite{EBIOS} detail the main steps that every organization should implement to deal with cyber risk.
However, they leave to the assessor the duty to decide how to quantify the risk by leveraging on security frameworks (e.g., ISO/IEC 27002 \cite{ISO27002} and NIST CSF \cite{NIST800-100}).
%
Such frameworks can be used in combination with different models of risk quantification.
They range from the simple \emph{two-factor} risk model that considers the risk as the product of the \emph{likelihood} that an asset is compromised times the \emph{impact} of the compromising, to the \emph{multi-factor} risk model where both likelihood and impact can be further decomposed into more detailed factors quantified independently and then aggregated. 
One of these latter models is the OWASP risk rating methodology \cite{OWASPRR}, in which the likelihood is estimated by considering 4 factors: threat agent, vulnerabilities, technical impact, and business impact.
This rating method is quite simple to adopt and OWASP provides guidelines about how to score each factor. However, the scoring is done by the human who needs to spend a big effort in the analysis and in remaining objective and consistent during it.
%
More objective evidences, coming from the infrastructure setup, should be considered to address such issues. A suitable approach to do it is leveraging an attack graph.
An attack graph allows representing possible ways via which an attacker can intrude into a target network by exploiting a series of vulnerabilities on various hosts. 
Attack graphs can be computed automatically starting from the data gathered in the system and thus they represent an objective model over which the risk may be computed~\cite{kaynar2016taxonomy}.
A large body of literature exists about using attack graphs to compute network security evaluation metrics (\cite{10.1145/1179494.1179502},\cite{6529081}), perform security risk analysis (\cite{10.1007/978-3-319-10329-7_3},\cite{GranadilloDMGAM18}) and compute near-optimal proactive defense measures (\cite{WangAJ14},\cite{GranadilloDMGAM18}).
In this scenario, Gonzalez et al.~\cite{GranadilloDMGAM18} propose an approach to perform data-driven cyber risk analysis through an attack graph, with a quantitative risk-aware system to dynamically manage risk response by considering likelihood, the induced impact, the cost of the possible responses, and the negative side-effects of a response.
However, the limitation of using the classical approaches to attack graphs is that almost all of them can support risk estimation by considering only technical aspects, without considering a comprehensive view of the organization environment (e.g., human factors, policies, and governance compliance).
There exist few proposals trying to address this limitation. Bonomi et al.\cite{Bonomi2020}
introduce a novel approach to attack graphs that is the multi-layer attack graph (MLAG). In the context of the PANACEA project~\cite{panacea} the architecture introduced by Gonzalez is enriched with comprehensive data through the MLAG.
Although these advancements represent a good starting point for comprehensive risk analysis, information about security governance and compliance is still not considered due to the difficulty to integrate them with technical data.
We faced such a challenge in~\cite{Angelini2020} by proposing a methodology towards a comprehensive risk analysis, including governance aspects. It consists of the review of a checklist-based cyber risk assessment, that we include in our methodology as detailed in Section~\ref{sec:ingr}.
\section{Problem Statement and Background Notions}
\label{sec:ingr}

Before describing our methodology, we introduce two models that are the inputs for our solution: a \emph{controls-based assessment} representing the governance perspective and a \emph{multi-layer attack graph} (MLAG) handling the technical perspective.
We define as \emph{controls-based assessment} a list of security controls and their individual evaluation, as for example the outcome of an assessment performed using ISO 27001~\cite{ISO27001}. A security control is everything that may require the definition and deployment of a specific process, policy, device, practice, or other actions which may contribute to the definition and quantification of risk. 
The evaluation for us is performed by assessing the coverage level of each control. Without loss of generality, we will assume a qualitative evaluation based on three levels i.e., completely covered (C), partially covered (PC), or not covered at all (NC) and it can be done by the assessor using data collected through interviews, workshops or questionnaires.
Let us note that compiling a controls-based assessment is a complex activity and prone to human bias because of the interpretability of several controls. Thus, our methodology proposes to reduce such bias.
Concerning the MLAG, we will consider the one presented in~\cite{Bonomi2020}.
This model allows to create an objective representation of the possible paths that an attacker may follow to compromise an asset and their likelihood can be quantified in an unbiased way depending on the severity of the vulnerabilities involved.
The MLAG is based on three interconnected layers: (i) the \emph{human layer} modeling how an attacker can compromise individual identities by exploiting human vulnerabilities of the personnel and  their relationships; (ii) the \emph{network layer} modeling the ICT part of the company used by individuals represented in the human layer; (iii) the \emph{access layer} modeling the credentials that humans may use to access devices (residing in the network layer).
Each layer is a multi-graph, meaning that multiple edges between the same pair of nodes are characterized by different vulnerabilities.

Using these models, we initially made a step towards their integration in~\cite{Angelini2020}, in which the MLAG is used to validate the controls-based assessment performed by a security expert, allowing to detect errors due to human bias.
Beyond the fact that the approach was completely manual, it just informs about the trustworthiness of a controls-based assessment.
Instead, the methodology we propose in this paper aims to quantify and re-asses the cyber exposure using comprehensive data and indicating the specific parts of the infrastructure they impact on, as well as to automate the computation steps.
\section{Methodology}
\label{sec:methodology}

In this section we present the methodology by firstly introducing the general description and then detailing the application of each activity.

\subsection{Methodology description}

\begin{figure}[ht!]
	\centering
	\includegraphics[width=0.9\linewidth]{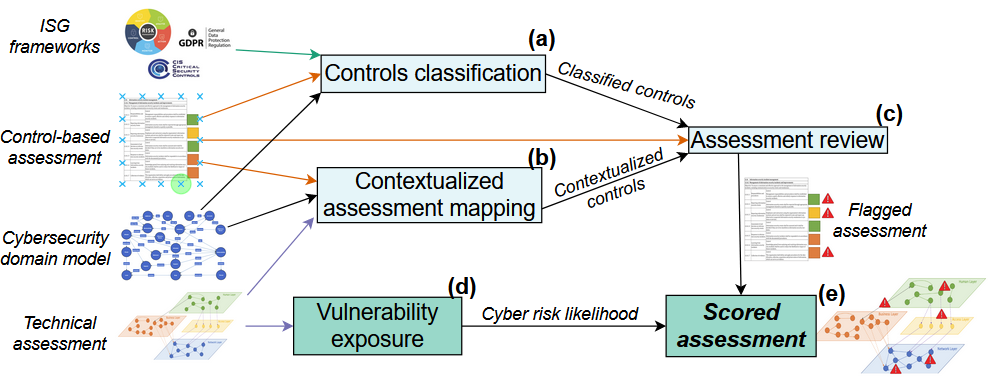}
	\caption{Scored assessment review methodology.\label{fig:methodology}}
\end{figure}

The proposed methodology (see Figure~\ref{fig:methodology}) starts taking as input (i) the reference frameworks, (ii) a controls-based assessment, (iii) domain models (either extracted from assessment and attack graph or modeled from scratch), and (iv) a technical assessment (through a MLAG).

%
\emph{Controls classification} (Fig~\ref{fig:methodology}a) is the first activity we must complete, which is labeling each security control and categorizing it according to a taxonomy that will allow us to map it to specific elements in the environment. A classified set of controls will be the result of this step.
This task is needed as security controls describe
sets of activities to be implemented to secure the organization and are expressed using different levels of granularity. 
%

\emph{Contextualized assessment mapping} (Fig~\ref{fig:methodology}b) is the second phase, in which we map security controls over the technical assessment result 
to identify the layers that each control affects.
Controls are mapped in one or more layers of the attack graph to accomplish this.
The result is a classification of security controls based on their context (that could be human, access, and/or network perspective). We denote such classification as contextualized controls.

\emph{Assessment review} (Fig~\ref{fig:methodology}c) is performed once we have the classified and contextualized controls, and it assigns a reliability evaluation to the assessment carried out for each control.
This task will result in a flagged assessment i.e., a controls-based assessment with additional information indicating the subset of controls that should be revised, further investigated,
or could be implemented involving different aspects.
This first part of the methodology represents the evaluation of governance and compliance aspects.

\emph{Vulnerability exposure} (Fig~\ref{fig:methodology}d) is concurrently assessed by using the MLAG and analyzing attack paths from human to network layer, using an approach inspired by Gonzalez~\cite{GranadilloDMGAM18}.
Finally, we map controls over the edges of the attack graph 
to obtain, for each edge, information about technical vulnerabilities, security controls, and assessed risk impacting specific parts of the organization.
After putting this information together, a \emph{scored risk assessment} (Fig~\ref{fig:methodology}e) is produced, computed with comprehensive and more objective information.

\subsection{Methodology application}

\subsubsection{Controls classification.}
Given the generality of the security controls, the first step of the methodology consists of classifying controls based on their features to assess how much they are clear to be interpreted for implementation.
Some controls are more focused because they refer to specific aspects of the organization (e.g., login and logout of devices), while others are more general and impact on different parts of the infrastructure (e.g., design of access control policies impact both the administrative Board and the ICT department).
This information will affect the evaluation of compliance with governance frameworks. General controls are more difficult to fully comply with due to the large number of concepts they include.
We consider two criteria to assess these aspects.
The first is the \emph{lifetime}, which is the point of the risk management process in which the controls must be put in place.
It specifies whether the control is implemented at \emph{design time} or \emph{run time}.
We assume that design-time controls have a greater impact on the risk management process because decisions made at design time impact also their implementation at run time as appearing in controls-based frameworks (\cite{ISO27001},\cite{NIST800-100}).
The second criterion is the \emph{management level} as it is introduced by Von Solms~\cite{VonSolms}: it indicates whether the nature of the security controls is \emph{operational}, meaning that they include practical activities, or \emph{compliance}, related to administrative responsibilities.

 We use two ontologies to classify lifetime and management levels of each control: one containing the semantics of security controls inspired by UCO~\cite{uco} and the other containing relations between lifetime operations (e.g., process execution monitoring, communication plans policies) and management levels.
 Then, we use ontology alignment between them using the automated tool AML~\cite{AML}. Ontology alignment is the process of determining correspondences between concepts in different ontologies~\cite{li2008rimom}: for each couple of concepts, such a process outputs a value, which describes the similarity of two terms of the input ontologies. Such a value ranges from 0 (no similarity) to 1 (full similarity).
 We compare the concept of each control to the concept of each feature (run time, design time, operational, and compliance), obtaining an output as shown in table~\ref{tab:management-example}.



\begin{table}[h!]
\centering
\begin{tabular}{lcccc}
\hline
\textbf{ID} & \textbf{Run time} & \textbf{Design time} & \textbf{Operational} & \textbf{Compliance} \\ \hline
A.9.4.3     & 0                & 0.923                & 0.926                 & 0                   \\ \hline
\end{tabular}
\caption{Example of alignment values for specificity degree.\label{tab:management-example}}
\end{table}

The classification of the controls labels the lifetime class with the label associated to the highest value between run time and design time alignments, while the management level class with the label of the highest alignment between operational and compliance.
If the alignment produces equal values for both run time and design time, or for both operational and compliance, the related class is labeled``Not defined'', indicating that lifetime or management level cannot be distinguished unambiguously.
Considering such classification, we define the \emph{specificity degree} of a control ($spec(c)$) as a function mapping the classification into numerical values.
The specificity degree is higher for run time and operational controls, while it decreases for design time and compliance ones. This is according to the assumption that run time and operational controls are easier to comply with because they are more focused on practical activities and more clear to interpret, while compliance and design tasks are harder to meet due to their generality.
Thus, we set a maximum value $\alpha$ for run time and operational controls, then we halve such a value each time that the feature is design time or compliance,
representing that more general is the control, more effort is required to comply with it.
In the ``Not defined'' case we do not decrease $\alpha$ because the related control has both operational/run-time and compliance/design-time nature.
The specificity degree values based on controls classification are reported in Table~\ref{tab:management-values}.

\begin{table}[h!]
\centering
\begin{tabular}{ll|lll|}
\cline{3-5} &  & \multicolumn{3}{c|}{Lifetime}   \\
\cline{3-5} &  & \multicolumn{1}{l|}{Runtime} & \multicolumn{1}{l|}{DesingTime} & Not defined \\
\hline
\multicolumn{1}{|c|}{\multirow{3}{*}{Management}} & Operational & \multicolumn{1}{c|}{$\alpha$}  & \multicolumn{1}{c|}{$\alpha/2$} & \multicolumn{1}{c|}{$\alpha/2$} \\ 
\cline{2-5} 
\multicolumn{1}{|c|}{} & Compliance  & \multicolumn{1}{c|}{$\alpha/2$} & \multicolumn{1}{c|}{$\alpha/4$}  & \multicolumn{1}{c|}{$\alpha/4$}\\ 
\cline{2-5} 
\multicolumn{1}{|c|}{} & Not defined & \multicolumn{1}{c|}{$\alpha/2$} & \multicolumn{1}{c|}{$\alpha/4$} & \multicolumn{1}{c|}{$\alpha/4$} \\ 
\hline
\end{tabular}
\caption{Values of specificity degree.\label{tab:management-values}}
\end{table}

The parameter $\alpha$ is the percentage of maximum simplicity of interpretation of the most focused controls and it is assignable by the assessor depending on the overall generality of the controls.
For example, a possible value we can assign to $\alpha$ is 0.5, indicating that at least half of the controls' activities can be completely complied with.
Considering such a value for the example in Table~\ref{tab:management-example}, the specificity degree for control A.9.4.3 is $0.25$ since it is classified as design time and operational.
\subsubsection{Contextualized assessment mapping.}
This step consists of mapping the controls-based assessment onto the layers of the attack graph to contextualize each security control to the infrastructure.
This is achieved with automatic ontology alignment~\cite{AML} between the ontology of security controls previously designed for controls classification and the ontological domain of the attack graph obtained by associating the nodes of the attack graph to ontology classes and the edges to ontology relations.
In this way, each control has an alignment value between 0 (not fitting) and 1 (full fitting) for each layer (human, access, network), as reported in the example of Table~\ref{tab:context-example}.

\begin{table}[h!]
\centering
\begin{tabular}{lccc}
\hline
\textbf{ID} & \textbf{Human} & \textbf{Access} & \textbf{Network} \\ \hline
A.9.4.3     & 0.7           & 0.371           & 0                \\ \hline
\end{tabular}
\caption{Example of alignment between controls and layers.\label{tab:context-example}}
\end{table}

The highest mapping value of each control represents the layer in which it is mainly contextualized. 
It is possible for a security control to be mapped on none of the layers, on one specific layer, or on multiple layers.
Controls that are mapped in a single layer are focused within the infrastructure of the organization (i.e., easier to comply due to homogeneity of the impacted resources), while the ones mapped in multiple layers impact different parts of the infrastructure
(i.e., more effort is required to comply).

Considering this rationale, we define the \emph{fitting degree} of a control ($fit(c)$) to express how much it fits the infrastructure.
Let $h_i$, $a_i$, $n_i$ be the alignment values for the control $c_i$ in the human, access and network layers respectively.
Let $l_M=max(h_i,a_i,n_i)$ be the highest between these values (most fitting layer) and $l_m=min(h_i,a_i,n_i)$ the smallest one (less fitting layer).
We define the fitting degree of a control as $l_M-l_m$, that is the range of the alignment values representing the spread of the control in the infrastructure.
Indeed, the fitting degree decreases as the control is more spread between layers. 
Considering the example in Table~\ref{tab:context-example}, the most fitting layer for control A.9.4.3 is human (with value $0.7$), while the layer in which it fits less is network (with value 0).
\subsubsection{Assessment review.}
Assessment review is performed following the approach we introduced in~\cite{Angelini2020}, which expresses the reliability of a controls-based assessment performed by a cybersecurity expert, in which each control can be assessed totally (C-Covered), partially (PC-Partially Covered), or nothing (NC-Not Covered).
In such an approach the security controls are associated with the layers of the attack graph and to management levels. Then the retrieved information is averaged to output a percentage of how much the performed assessment is reliable for each security control.
In particular, it is computed the average of the following four data: lifetime, management level, mapping value of a control in the layers, and the coverage factor of the assessment.
While the calculation of these data is based on manual quantification, we extend it by automatizing the computation of these values through ontology alignment using the ontologies introduced in the previous sections.
We define the output we obtain by such a computation as \emph{reliability degree} of a control ($rel(c)$), which represents the trustworthiness of the assessment of each control.

\subsubsection{Vulnerability exposure.}
We use the approach of Gonzalez et al.~\cite{GranadilloDMGAM18} to evaluate the probability that an attack step over the MLAG is performed, indicating the difficulty to exploit the related vulnerability.
They model each attack path as a Markov chain and then evaluate the exit rates ($\lambda$) of each node as the sojourn time in each state: such a rate is the likelihood that the vulnerability in the related edge of the attack graph is exploited.
Such exit rates are evaluated based on the attributes of each vulnerability, therefore we need to distinguish the network layer from the human/access ones, because in the former such information is available through CVSS~\cite{CVSS}, while in the latter we retrieve human vulnerabilities from ISO 27005~\cite{ISO27005}.
Then the qualitative values of each attribute are normalized in the range [0,1] such that the higher the values, the higher the severity.

In the network layer we can retrieve the following metrics valued according to CVSS: Attack Complexity (AC)
, Attack Vector (AV)
, Privilege Required (PR)
, Exploit Code Maturity (CM) 
and Report Confidence (RC). 
The attacker ability must be considered for each attribute and it can be \{Naive, Advanced, Professional\}.
Thus, given a network vulnerability $v_k$ of the k-th step of the attack path, assuming attacker with abilities $A =<t_{AC}, t_{AV}, t_{PR}, t_{CM}, t_{RC}>$ and considering the attributes of network vulnerability as $X(v_k)$, the difficulty of exploitation of the edge $e_k$ in the network layer is:
\[
\lambda_{k} = \prod_{X \in {AC,AV,PR,CM,RC}}H(X(v_{k})-t_{X})*X(v_{k})
\]
where $H(z)$ is the Heaviside step function equal to 0 if z$<$0 and 1 otherwise.

In human and access layers, we refer to vulnerabilities and their evaluation guidelines contained in ISO 27005, as ``No logout'', ``Sharing credentials'', ``E-mail misuse''.
We consider the attack complexity (AC) and the access vector (AV) attributes of each vulnerability, such that AC can assume values \{Low, High\} and AV can assume values \{Proximity, Knowledge\}.
We do not consider the attacker ability because the vulnerability exposures in these layers do not depend from the attacker. 
Following this approach, given a vulnerability $v_k$ related to the k-th step of an attack path, with attributes $X(v_k)$, then:
\[
\lambda_{k} = \prod_{X \in {AC,AV}}X(v_{k})
\]

\subsubsection{Scored assessment.}
At this point, we have the necessary data to compute a \emph{comprehensive score} for each edge of the attack graph including both governance compliance and technical aspects.
On one side we have information about controls (i.e., specificity degree ($spec(c)$), fitting degree ($fit(c)$), reliability degree ($rel(c)$)) and their classification in the attack graph layers based on the information computed in the contextualized assessment mapping phase of the methodology.
We use such information to introduce a \emph{governance factor} of a layer expressing the degree of compliance of the layer with respect to security governance frameworks (i.e., security controls).
To do so, we group the controls according to the layers they belong to.
Let $C_i$ be the set of controls mapped into layer $l_i$, then the governance factor is:

\[
governanceFactor(l_i) = gov(l_i) = avg\{ \forall c \in C_i [f(spec(c),fit(c),rel(c))]\}
\]

Function \emph{f} is a generic aggregation function that maps the relation between specificity, fitting, and reliability degrees into a value between 0 and 1.
Suitable solutions for such a function could be the average to assign to the three values the same importance, as well as the minimum (worst-case scenario usually considered by cyber-risk assessment and management activities).
Once aggregated,
the average between the compliance degree of all the controls belonging to the same layer is computed: this allows to consider the mean compliance of the governance within each layer.

On the other hand we know the difficulty of technical exploitability of each vulnerability through the computation of $\lambda$ rates associated with the edges of the attack graph.
The combination of the governance factor with $\lambda$ values represents the association of governance and technical knowledge.
We associate the governance factor of a layer to all the edges belonging to that layer: this indicates an equal impact of that factor on each edge.
Considering this assumption, we define the \emph{comprehensive score} in the edge $e_{k}$ in layer $l_{i}$ of the attack graph as:

\[
comprehensiveScore(e_k) = f(gov(l_i), \lambda_k)*cv
\]

where $cv$ is a constant set by the security expert that takes into account how rigorous is the assessment evaluation, and it weights the comprehensive score. It is defined as:
\[
cv = \frac{\sum_{x\in X}\alpha_x*|x|}{|X|}
\]
where X is the set of possible values used to assess the security controls
and $|X|$ is the number of such different values (i.e., 3 if we consider C, PC, and NC); $\alpha_x$ is the percentage of coverage related to each assessment value (e.g., 
$\alpha_{PC}=0.5$ means that the assessor evaluates "PC" all the controls in which 50\% of the activities have been implemented); $|x|$ is the number of controls that have been assessed with value ``x''.

The aggregation function \emph{f} in the \emph{comprehensive score} calculation can be chosen depending on the relation between the compliance governance and the technical exposure (e.g., average, min, max).
The comprehensive scores projected onto the technical assessment represent the cyber exposure of the infrastructure in a more fitting way.
\section{Validation}
\label{sec:val}
We illustrate a case study to evaluate a realistic implementation of the methodology and its behavior first in a controlled fitting assessment and then in two controlled biased assessments.
Then, we relax the hypothesis of controlled bias by simulating possible scenarios of random human errors. Finally, we test limit cases to study the bounds of the methodology.
\subsection{Case study: controlled assessment ISO/IEC 27001:2013}
\begin{figure}[ht!]
	\centering
	\includegraphics[width=0.8\linewidth]{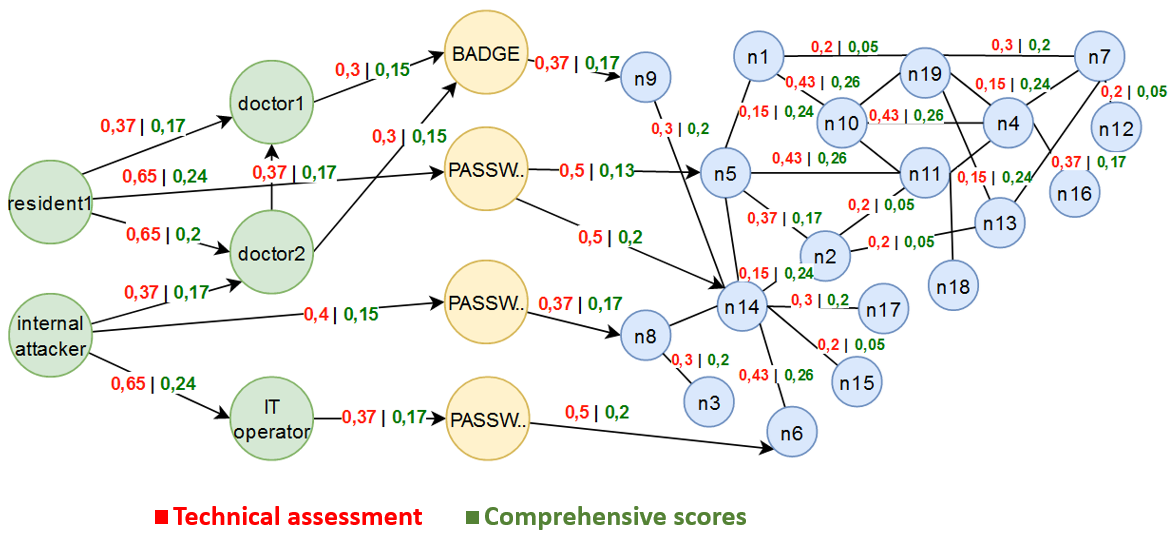}
	\caption{Output of scored assessment review methodology.\label{fig:mlag}}
\end{figure}

\begin{figure}
    \centering
    \subfloat[\centering]{{\includegraphics[width=0.4\linewidth]{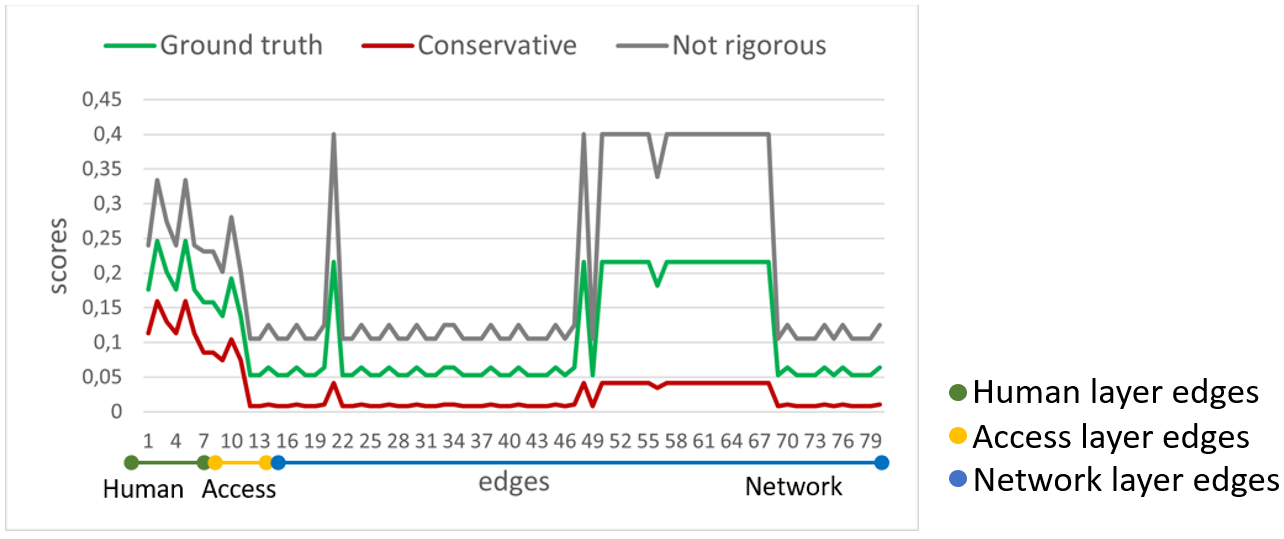} }}
    \qquad
    \subfloat[\centering]{{\includegraphics[width=0.4\linewidth]{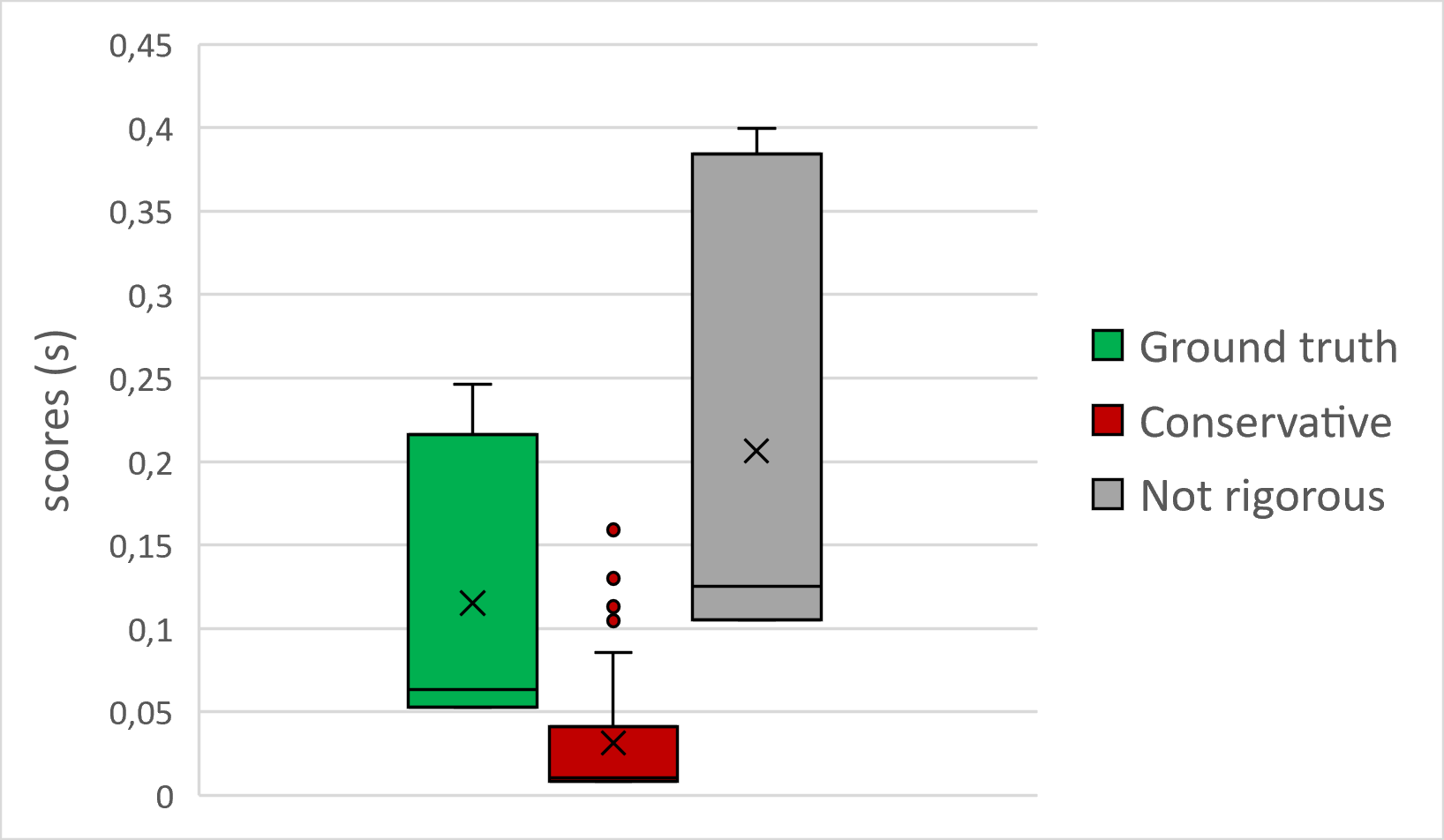}}}
    \caption{Comparative charts of biased assessments with ground truth.}
    \label{fig:two-charts}
\end{figure}

We consider a healthcare organization (hospital) whose network is reproduced in the PANACEA emulation environment~\cite{panacea}, while the assessment is simulated to capture a realistic scenario.
The infrastructure is represented in Fig.~\ref{fig:mlag} through the MLAG.
The human layer is reported in green in the left part of the figure, composed of the IT department, a resident, two doctors, and an employee simulating a possible internal attacker.
The yellow nodes in the middle part of the graph represent the access layer in which everyone accesses the systems through username and password and \emph{doctor1} has to use a badge to access the device \emph{n9}.
Blue nodes in the right part of the graph compose the network layer which is a dense network composed of physical devices and several routers.\\
The MLAG contains a set of vulnerabilities from CVE~\cite{CVE}, namely: 2010-1883, 2019-2018, 2009-2412, 2018-4846, 2018-3110.
The vulnerabilities in human and access layers are the following: (i) sharing credential (i.e., the person shares the personal credentials with others); (ii) no log-out when leaving the workstation (i.e., the person leaves unattended device(s) logged in with personal credential); (iii) poor password management (i.e., the person uses weak passwords and do not update them).
%
The assessment has been conducted using ISO/IEC 27001:2013 controls because it is common for the organization to be certified by ISO and it is such that higher assessment scores correspond to a lower cyber risk.
We label the fittest controls-based assessment as \emph{ground truth} and
in Fig.~\ref{fig:mlag} the assessment values before the review (i.e., considering only technical perspective) are reported in red, and the comprehensive scores after the review are shown in green.
The comprehensive scores are lower than the technical assessment in 94\% of the edges with a range of variation between 10-30\% (mean 25\%).
This highlights that while the technical assessment would indicate a certain level of cyber risk due to the presence of specific vulnerabilities, this is not correct. In fact the comprehensive scores (that consider governance factor across multiple layers) indicate that the organization is overall not compliant with the standards, resulting in higher cyber exposure. Considering just the technical perspective would give a distorted overview of the cyber risk.\\
Then we introduce two controlled biased assessments to study the behavior for the following cases:
(i) the \emph{conservative assessment} represents the assessment performed by an expert that gives the certification in a meticulous way. The bias is that if something is only partially covered, the related controls are flagged as ``not covered'';
(ii) the \emph{not rigorous assessment} represents the opposite case: it simulates the assessment biased by an auditor more oriented to flag partially covered controls as ``covered''.
We perform the methodology for the assessments using the average as aggregation function for \emph{governance factor} and \emph{comprehensive score}.
We obtain as result the distribution of the scores reported in Fig.~\ref{fig:two-charts}: the ground truth has a mean score equals to 0.115 and a standard deviation of 0.0746; the conservative assessment has a mean of 0.0314 and a standard deviation of 0.036; the not rigorous assessment has a mean of 0.206 and a standard deviation of 0.125.
The trend is the same in all the cases; however, they differ for the values the scores assume:
the box plots in Fig.~\ref{fig:two-charts}b make clear the difference with the ground truth.
The conservative assessment has some outliers due to the high number of ``NC'' that appears with respect to the rest of the assessment. In the not rigorous assessment, the great amplitude of the box underlines the underestimation of cyber exposures.
A big difference with respect to the ground truth can be dangerous: it either means that the risk coverage is not sufficient, therefore several outliers appear (conservative), or that the risk is underestimated, therefore the distribution of scores is amplified (non-rigorous).
This information (i.e., presence of outliers and amplified distribution) can be used to detect which parts of the infrastructure have overestimated or underestimated exposures by considering the line-chart of Fig.~\ref{fig:two-charts}a which reports the edges IDs in the x-axis and the comprehensive scores in the y-axis.
\subsection{Assessment review sensitivity analysis.}
To test the methodology sensitivity we relax the hypothesis of controlled bias by perturbing the ground truth assessment of a certain percentage of error and studying how this perturbation affects the methodology scores.
We tested 28 cases: seven cases for each percentage of error equal to $15\%$, $45\%$, $65\%$, and $90\%$. These values are chosen by considering experimental results that help in differentiating the different cases of errors.
\begin{figure}[ht!]
	\centering
	\includegraphics[width=0.8\linewidth, height=5cm]{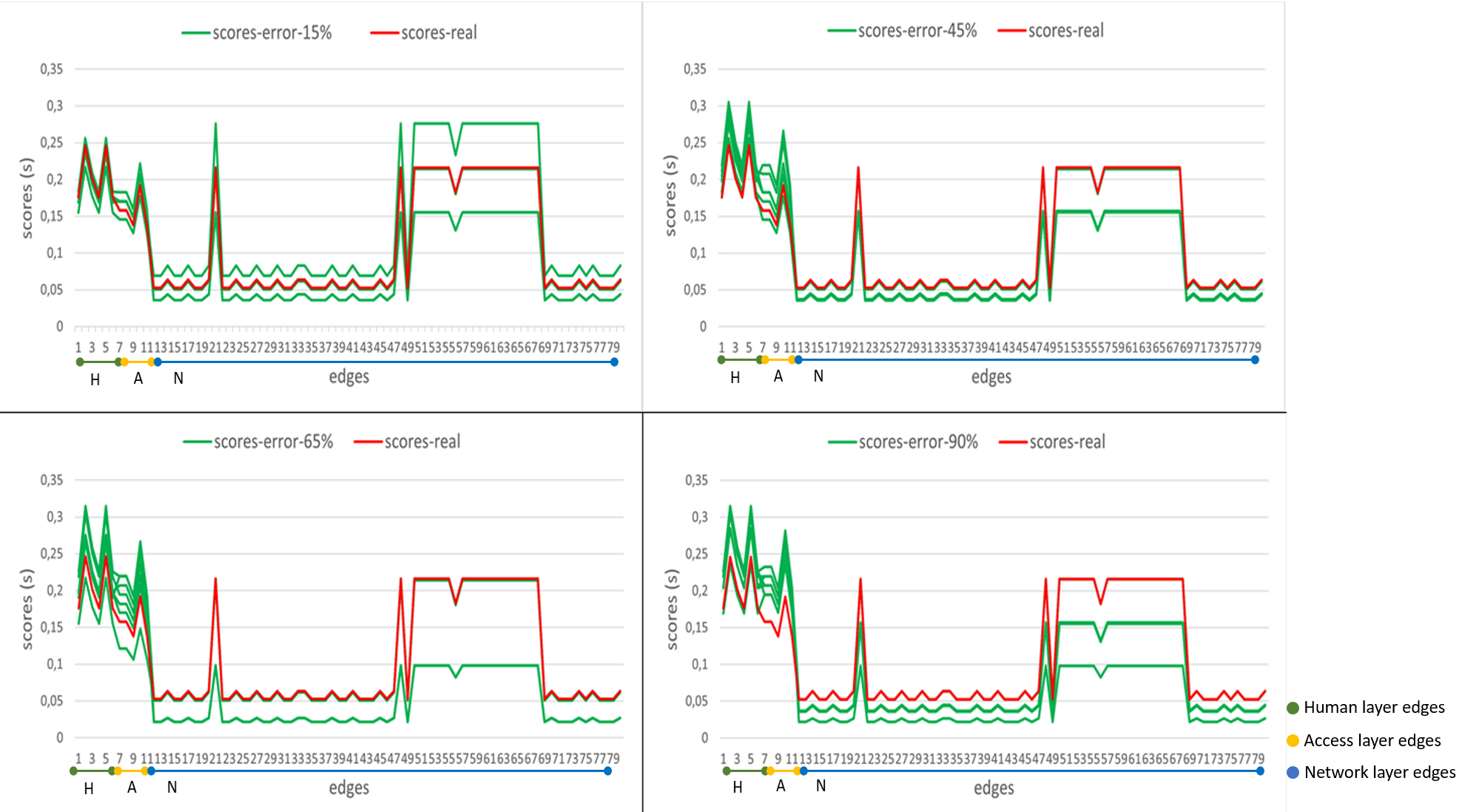}
	\caption{Distribution of the 28 error cases and ground truth.\label{fig:err}}
\end{figure}
Fig.~\ref{fig:err} represents the trend of the different cases.
We notice that the scores affected by errors are more variable for human and access layers (the first ones in the x-axis): this is because in such layers we have fewer edges, therefore the information about controls is more compacted in each edge and this increases the variability. Instead in the network layer it is more distributed among the edges, therefore the assessment scores are more uniform.
Also, we can see how the variation of the errors (in green) differs from the ground truth (in red) in the four cases. 
In the human and access layers the errors are detectable because they tend to amplify the value of the scores, while in the network layer (last edges in the x-axis) the errors are detectable because they tend to decrease the values of the scores.
This means that it is possible to detect the presence of error with respect to the ground truth in the vast majority of cases by analyzing the peaks. 
Thus the methodology is able to distinguish the degree of bias in the perturbed assessments from the unbiased one (ground truth).
\subsection{Scores distribution validation.}
\begin{figure}
\centering
\begin{minipage}{.5\textwidth}
  \centering
  \includegraphics[width=\linewidth,height=3cm]{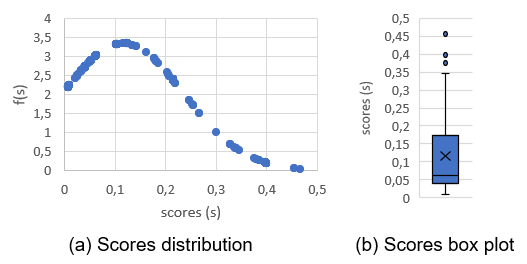}
  \captionof{figure}{Scores trend.\label{fig:general-analysis}}
\end{minipage}%
\begin{minipage}{.5\textwidth}
  \centering
  \includegraphics[width=0.5\linewidth,height=3cm]{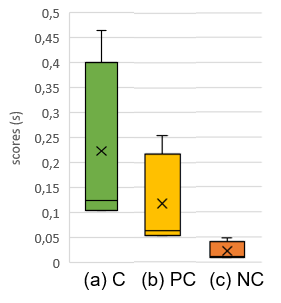}
  \captionof{figure}{Box plots of borderline cases. \label{fig:border-boxplot}}
  \label{fig:test2}
\end{minipage}
\end{figure}

We finally remove the hypothesis of perturbed biased assessment by studying the scores the methodology produces independently from the context. It is useful to ensure adaptability to a generic context and it allows to understand the bounds of the scores variability. The data are obtained firstly by considering all the possible combinations of assessment (Fig.~\ref{fig:general-analysis}) for 114 security controls.
Then, we considered the three borderline cases, where the controls are all covered, partially covered, or all not covered (Fig.~\ref{fig:border-boxplot}).

In Fig.~\ref{fig:general-analysis} the Gaussian distribution (\ref{fig:general-analysis}a) and the box plot (\ref{fig:general-analysis}b) of the scores are reported.
They have a mean of 0.117 and a standard deviation of 0.120, with distribution concentrated in the part of the chart with lower values.
The outliers are all located in correspondence with high values: they can be considered the more secure points of the infrastructure.
The presence of a few of those points indicates that the methodology is oriented to identify the vulnerable parts of the organization, and this is in line with expectations due to the applicability of this approach to critical infrastructures.
%
%
%
In Fig.~\ref{fig:border-boxplot} the three borderline cases of the possible bias are reported.
For the assessment with all controls covered (Fig.~\ref{fig:border-boxplot}a) the distribution is more amplified than the other cases because several security operations are implemented, therefore the organization is compliant from both governance and technical perspectives and the scores assume higher values. For some elements they are still low because some controls deal with critical parts of the organization or they are too general and difficult to be compliant with (e.g., managing changes to supplier services is a control that depends also on third parties). Also, we set the parameter at the minimum (e.g., $\alpha=0.5$) because we are considering a critical infrastructure, therefore we tend to be conservative.
In the other cases (Fig.~\ref{fig:border-boxplot}b,\ref{fig:border-boxplot}c) the distribution is similar to the case of full coverage, but the mean is halved.
This is a good result because it indicates that the methodology takes the assessment variation into account.
Also, the standard deviation is 0.1 in case (a), 0.078 in case (b), and 0.016 in case (c), indicating good stability of the computed scores independently from the context.
\section{Conclusion}
\label{sec:con}

In this paper we proposed a methodology for obtaining a more fit risk assessment of a generic organization exploiting a multi-layer attack graph. The methodology has been validated through a case study and statistical tests, proving its capability to correct cyber risk underestimation due to human assessor errors or biases. 
Some limitations can be identified, that will be the object of further research: the aggregation functions between governance and technical factors could be more complex than the average we used for the validation, as well as the ontology models for security controls and management levels could be not exhaustive in terms of modeled information.
Indeed, we plan as future work (i) to provide an aggregation function able to capture with suitable weights the governance and technical aspects and the refinement of the proposed ontologies; (ii) to develop a Visual Analytics~\cite{Keim2008} tool for the proposed approach, to better support the application of this methodology. 
%
%
%
\bibliographystyle{splncs04}
\bibliography{mybibliography}

\end{document}